\documentclass[preprint,12pt]{elsarticle}

\usepackage{amssymb}

\journal{Nuclear Instruments and Methods in Physics Research A}

\begin{document}

\begin{frontmatter}

\title{A reconstruction method for neutrino induced muon tracks 
taking into account the apriori knowledge of the neutrino source}

\author{A.G. Tsirigotis\corref{cor1}}
\cortext[cor1]{Corresponding author. Tel: +302610367517; Fax: +302610367528}
\ead{tsirigotis@eap.gr}
\author{A. Leisos}
\author{S. E. Tzamarias}
\address{Physics Laboratory, School of Science \& Technology, Hellenic Open University}
\author{On behalf of the KM3NeT Consortium}

\begin{abstract}
Gamma ray earthbound and satellite experiments have discovered, over the last years, many galactic and
extra-galactic gamma ray sources. The detection of astrophysical neutrinos emitted by the same sources
would imply that these astrophysical objects are charged cosmic ray accelerators and help to resolve the
enigma of the origin of cosmic rays. A very large volume neutrino telescope might be able to detect these
potential neutrino emitters. The apriori known direction of the neutrino source can be used to effectively
suppress the $^{40}K$ optical background and increase significantly the tracking efficiency 
through causality filters. We report on advancing filtering and
prefit techniques using the known neutrino source direction and first results are presented.
\end{abstract}


\end{frontmatter}

\section{Introduction}
\label{intro}
The future Mediterranean very large volume neutrino telescope, KM3NeT, will be one of the world's 
largest particle detectors and will provide a research infrastructure for a rich and 
diverse deep-sea scientific program \cite{km3netcdr,km3nettdr}.
Various astrophysical sources are expected to produce high-energy neutrinos that 
can be detected with KM3NeT. 
The most promising candidate neutrino sources are Galactic point-like 
(or with a small angular size) sources.
The observation of neutrinos emanating from such sources would bring unique new insights
on the nature of cosmic accelerators and might help to resolve the enigma of the origin of cosmic rays.
Observations by gamma ray earthbound and satellite experiments have revealed 
many such astrophysical objects, in which high-energy processes at and beyond the TeV energy level take place. 
However, measurements with gamma rays alone cannot clearly distinguish whether the 
accelerated particles are leptons or hadrons. 
Only the observation of neutrinos from a source can unambiguously establish the 
hadronic nature of that source.

However, the Galactic sources are generally expected to have a cut-off in their energy spectra 
in the range $1-10$~TeV. The signal produced by neutrinos in this energy range could be lost among
the $^{40}K$ optical background, resulting in low track reconstruction efficiency.
In this work we report on advancing triggering, filtering and prefit techniques that use the apriori 
known direction of the neutrino source to effectively
suppress the $^{40}K$ optical background and enhance the detection efficiency for neutrinos 
from Galactic sources for which the direction is known from their gamma ray emission.
\section{Detector description and simulation framework}
\label{simdet}
In the present study the telescope layout considered is the one optimized during the KM3NeT Design Study \cite{km3nettdr}, 
exhibiting optimal sensitivity in discovering astrophysical point sources emitting neutrinos with an
energy spectrum of $E^{-2}$ and either a high energy cut-off or no cut-off at all. 
According to this layout, the KM3NeT detector will consist of 12320 photo-sensors distributed 
over 308 Detection Units (DUs).
The photo-sensor unit is a digital optical module (DOM) consisting of a 17-inch diameter pressure resistant
glass sphere housing 31 3-inch photomultiplier (PMT) tubes \cite{DOM}.
A DU is a vertical structure which carries photo-sensors
and devices for calibration and environmental measurements, arranged vertically on
20 Storeys.
Each Storey consists of a bar with one DOM at either end. 
The horizontal distance between the centers of the DOMs is 6~m. The vertical distance between Storeys
is 40~m, while the position of the lowest Storey is 100~m above the seabed.
The bars have an orientation orthogonal to their neighbors.
The distribution of the positions of the DUs on the seabed (the so-called
footprint) is homogeneous. The footprint forms a roughly circular shape and has a typical DU
density corresponding to an average distance between neighboring detection units of about 180~m.
The total instrumented volume of the detector is 5.8~km$^3$.

For the present study we have simulated the detector signal produced by neutrinos using the HOURS 
(Hellenic Open University Reconstruction and Simulation) physics analysis package \cite{HOURS}.
We have used a generic neutrino flux distributed isotropically (in a $4\pi$ solid angle) on the Earth's atmosphere,
with an energy distribution following a power law spectrum, with a spectral index of $-2.0$, in the range of $15~\mathrm{GeV}-100~\mathrm{PeV}$. 

In the following, the signal produced by the secondaries of a neutrino interaction event in the vicinity of the neutrino telescope is described by a collection of active DOMs (hits), $h_i(t_i,\vec H_i,m_i)\ (i=1,..,N_{hit})$. Each hit is described by the first photon arrival time, $t_i$, the active DOM's position, $\vec H_i$, and the multiplicity, $m_i$. The hit multiplicity is the number of active 3-inch PMTs of a DOM within a time window of 10~ns. A level one (L1) coincidence trigger is defined to 
fire/be activated for $m_i>1$.

A random background rate of 5~kHz is assumed for each 3-inch PMT, including dark current, $^{40}K$ decays,
and bioluminescence. In addition to random coincidences, an L1 hit (with multiplicity $m=2$) rate of 500~Hz is assumed on each DOM, due to genuine coincidences from $^{40}K$ decays.

For each signal event, produced by a neutrino emanating from a known astrophysical point source, we assume that the direction vector, $\hat d$, of the muon track coincides with the source direction. 
In order to estimate the fake signal produced by implying in the analysis a certain direction, 
we assume for background events 
(atmospheric neutrinos and atmospheric muon bundles) a random (fake) candidate muon track direction 
distributed isotropically in a $4\pi$ solid angle.
\section{Background filtering}
\label{filtering}
The filtering technique of the $^{40}K$ optical background is based on a causality criterion 
that uses the assumed
direction vector, $\hat d$, of the muon track generated by the incident neutrino.
The arrival time, $t_i$, of a photon emitted by the muon with the Cherenkov angle, $\theta_c$, to a DOM located at position $\vec H_i$ satisfies the relation:
\begin{equation}
\label{eq1}
ct_i=a_i+b_itan\theta_c,
\end{equation}
where
$a_i=\hat d \cdot (\vec H_i - \vec V)$, and
$b_i=|\vec H_i - \vec V - a_i \hat d |$
is the vertical distance of the DOM to the muon track.
$\vec V$ is the position vector of the pseudo-vertex, 
corresponding to the position of the muon when the time measurement started.

Two hits $h_i(t_i,\vec H_i,m_i)$ and $h_j(t_j,\vec H_j,m_j)$ produced by direct photons on two different DOMs
according to Eq.~\ref{eq1} satisfy the following relation, in which $\vec{V}$ is eliminated:
\begin{equation}
\label{eq2}
\frac{c\Delta t - \hat d \cdot \Delta \vec H}{tan\theta_c}=\Delta b,
\end{equation}
where
$\Delta t=t_i-t_j$, $\Delta \vec H=\vec H_i-\vec H_j$ and $\Delta b=b_i-b_j$.
If we project the DOM's positions on a plane perpendicular to the assumed direction $\hat d$, 
then from simple geometry and using Eq.~\ref{eq2} we have the relation:
\begin{equation}
\label{eq3}
|\Delta b|=\left|\frac{c\Delta t - \hat d \cdot \Delta \vec H}{tan\theta_c}\right| < |\Delta \vec H-(\hat d \cdot \Delta \vec H)\hat d|.
\end{equation}
According to Eq.~\ref{eq3} we form the causality criterion to be satisfied by the two hits:
\begin{equation}
\label{eq4}
|c\Delta t - \hat d \cdot \Delta \vec H|<tan\theta_c |\Delta \vec H-(\hat d \cdot \Delta \vec H)\hat d|+ct_s,
\end{equation}
where $t_s=10~\mathrm{ns}$ is inserted in order to relax the causality requirement, 
due to light dispersion and time jitter effects.
In order to limit the number of noise hits satisfying this criterion, we also require the
longitudinal distance between the two DOMs along the direction of the muon track to be:
\begin{equation}
\label{eq5}
|\hat d \cdot \Delta \vec H|<800~\mathrm{m},
\end{equation}
while the corresponding lateral distance should be less than one absorption length\footnote{This is 
the maximum absorption length for wavelength 440~nm.}:
\begin{equation}
\label{eq6}
|\Delta \vec H-(\hat d \cdot \Delta \vec H)\hat d|<67.5~\mathrm{m}.
\end{equation}
Equations \ref{eq4}, \ref{eq5} and \ref{eq6} define the causality condition that has to be satisfied by any two hits.

In order to use this condition as criterion for background filtering, 
for every hit, $h_i$, we calculate the following sum:
\begin{eqnarray}
M_i=\sum_{j=1}^{N_{hit}}m_jC(h_i,h_j) \nonumber
\end{eqnarray}
where $m_j$ is the multiplicity of the hit $h_j$ defined in Section \ref{simdet},
while $C(h_i,h_j)=1$ if the causality condition is satisfied by the two hits, or 0 otherwise.
If $M_i<5$ the hit $h_i$ is rejected.
When we apply this filtering technique we achieve a 99.7\% rejection of the noise hits, 
while more than 90\% of the signal hits survive.
\section{Triggering and prefit}
\label{tripre}
For every three hits on different DOMs, that when combined in pairs satisfy the causality condition, 
a pseudo-vertex, $\vec V$, can be found 
analytically, as described in the following. Using Eq.~\ref{eq1} we calculate that:
\begin{eqnarray}
\label{eq7}
ct_i=\hat d \cdot (\vec H_i - \vec V)+b_i tan\theta_c \Rightarrow  \nonumber \\
c_i+z'=b_i \ (i=1..3),
\end{eqnarray}
where
\begin{equation}
\label{eq8}
z'=\frac{\hat d \cdot \vec V}{tan\theta_c}
\end{equation}
and $\displaystyle c_i=\frac{ct_i-\hat d \cdot \vec H_i}{tan\theta_c}$.\\
Then, we define a new coordinate system with the $z$-axis parallel to the assumed direction $\hat d$, 
the origin at the point defined by $\vec H_1$, 
and the $xz$-plane defined by the $z$-axis and the point defined by $\vec H_2$. 
The base vectors of this coordinate system are:
\begin{eqnarray}
\hat x&=&\displaystyle\frac{1}{p}(\vec H_2-\vec H_1) - \displaystyle\frac{1}{p}\left[\hat d\cdot(\vec H_2-\vec H_1)\right]\hat d \nonumber \\
\hat y&=&\displaystyle\frac{1}{p}\hat d \times (\vec H_2-\vec H_1) \nonumber \\
\hat z&=&\hat d \nonumber
\end{eqnarray}
In this frame, the projections of the three hits that we consider
on the $xy$-plane are: $\vec H_1(0,0),\ \vec H_2(p,0),\ \vec H_3(q,r)$, 
where the values of $p,~q$ and $r$ can be estimated easily from the positions of the hits.
The unknown pseudo-vertex $\vec V(x,y,z)$ is expressed as follows:
\begin{equation}
\label{eq9}
\vec V=\vec H_1 + x\hat x + y\hat y + z\hat z
\end{equation}
where from Eq.~\ref{eq8}:
\begin{equation}
\label{eq10}
z=\hat d \cdot (\vec V - \vec H_1)=z'tan\theta_c-\hat d \cdot \vec H_1
\end{equation}
Taking into account that $b_i$ in Eq.~\ref{eq7} is the vertical distance of the hit position from the track, 
then the $x,~y,~z'$ satisfy the following set of equations:
\begin{eqnarray}
b_1^2=(c_1+z')^2&=&x^2+y^2 \nonumber \\
b_2^2=(c_2+z')^2&=&(x-p)^2+y^2 \nonumber \\
b_3^2=(c_3+z')^2&=&(x-q)^2+(y-r)^2 \nonumber
\end{eqnarray}
with $p,r\neq 0$. The above set of equations and Eq.~\ref{eq10} can be used to calculate the explicit values of $x,y,z$, 
and then from Eq.~\ref{eq9} the pseudo-vertex.

The probability to find three L1 noise hits\footnote{Produced by PMT dark current, $^{40}K$ decays and bioluminescence. 
In Section \ref{fake} noise from atmospheric neutrinos and muons is examined.} producing a pseudo-vertex has been found in our simulation studies to be of the order of $10^{-5}$. 
Consequently, the requirement to find at least one pseudo-vertex using only the L1 hits of an event, offers a good directional selection criterion (trigger) when searching for muons arriving from a certain direction.

\begin{figure}
\begin{center}
\includegraphics[width=0.8\textwidth]{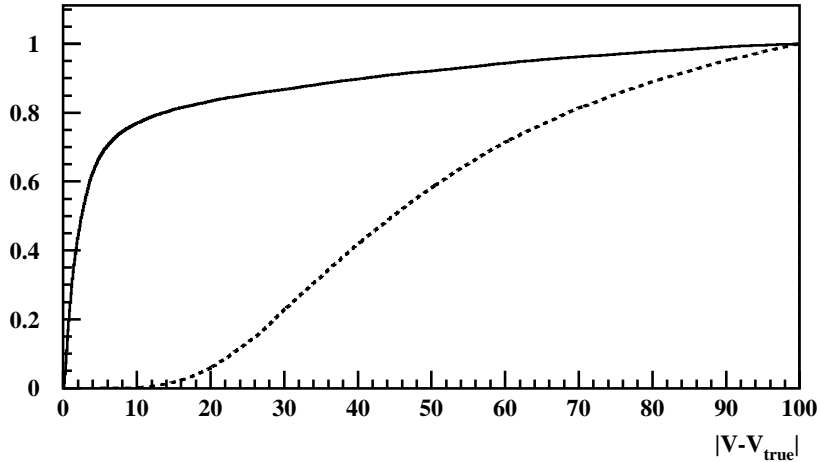}
\end{center}
\caption{The cumulative distribution of the distance between the estimated pseudo-vertex
and the true track
pseudo-vertex in meters (solid line). The median of the distribution is 2~m. 
The corresponding distribution for the alternative prefit technique used in HOURS, 
where an apriori known track direction is not used, is also shown (dashed line).}
\label{psfvertex}
\end{figure}

Obviously, in a signal event, the above procedure finds many candidate pseudo-vertexes using different hit triplets. 
Then we examine the distribution of these pseudo-vertexes in space searching for clusters. The 
cluster with the most participants, within a radius of 5~m, is chosen to estimate the position of the track pseudo-vertex.
This track pseudo-vertex is defined as the average position of the participants in the cluster. 
We have found, using the simulated neutrino events described in Section \ref{simdet}, that this prefit technique estimates 
the true track pseudo-vertex with an accuracy of 2~m, as it is presented in Fig.~\ref{psfvertex}.

The estimated track pseudo-vertex and the assumed direction, $\hat d$, can be used in order to reduce further the contribution of noise hits.
For this we require the time residual of the hits to be less than 150~ns ($|t_i-t_i^{exp}|<150~\mathrm{ns}$), 
where the expected time can be derived according to Eq.~\ref{eq1} when using for $\vec V$ the position of the estimated 
track pseudo-vertex.
Furthermore, we require that the distance of the selected hits from the estimated track (expressed by the vector $\hat d$ and the track pseudo-vertex) to be less than 4 light absorption lengths in the sea water ($\sim 270~\mathrm{m}$).
After this track prefit stage the 99.97\% of the noise hits have been rejected, 
while 90\% of the signal hits remain.
\section{Muon track reconstruction and results}
\label{reco}
\begin{figure}
\begin{center}
\includegraphics[width=0.8\textwidth]{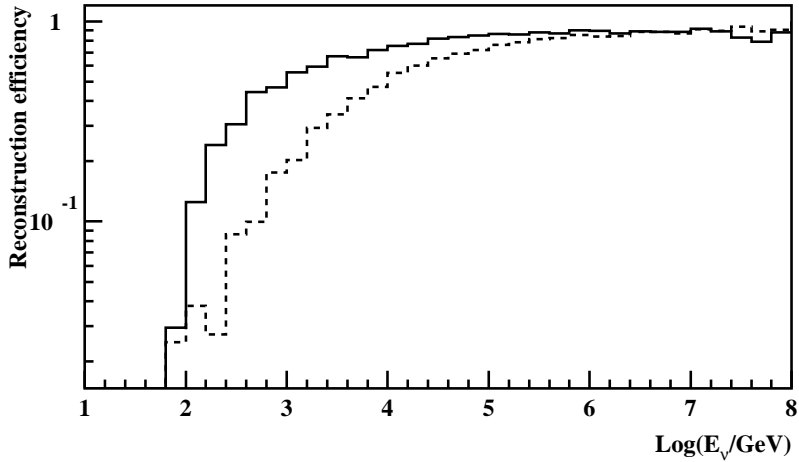}
\end{center}
\caption{Reconstruction efficiency versus neutrino energy for events with at least three signal L1 hits (solid line). 
The corresponding efficiency for the alternative fitting technique used in HOURS is also shown (dashed line).
The lower values of the solid line than those of the dashed line at ~7.5 are a statistics effect.}
\label{efficiency}
\end{figure}
We use a combination of a $\chi^2$ fit and a Kalman Filter selection algorithm in order to reconstruct muon tracks
as described in \cite{kalman}. 
The hits in the track reconstruction are those surviving the prefit stages described in Sections \ref{filtering} 
and \ref{tripre}. 
At this tracking stage the assumed direction $\hat d$ and the estimated track pseudo-vertex are used as 
initial values for the fit, however the track parameters are not constrained to the assumed direction.

This tracking algorithm produces many candidate tracks.
We select the best candidate track by employing information offered by the hit arrival times 
and Multi-PMT direction likelihood as described in \cite{kalman}.
Specifically, for each candidate track we evaluate the likelihood 
value, using the hits that have survived the background filtering stages.
The best candidate track, with the largest likelihood value, is further improved by performing a final likelihood fit.
However, the estimated track is finally accepted when it is misaligned with respect to the assumed direction vector, $\hat d$, by less than $1^\circ$\footnote{For point source searches this angular difference has to be further optimized.}.

In Fig.~\ref{efficiency} the reconstruction\footnote{The tracking quality criteria applied are minimal. 
We only require more than seven hits to participate in the best candidate solution found by the tracking algorithm.} efficiency versus the neutrino energy is shown for simulated events, containing
at least three signal L1 hits, produced as described in Section \ref{simdet}.
The corresponding efficiency of the alternative fitting procedure used in HOURS (when the apriori known track direction is not used at all) is also shown.
The reconstruction efficiency is improved
by a factor of 2.5 for 1~TeV neutrino induced events, and a factor of 1.5 for 10~TeV neutrino induced events.

\begin{figure}
\begin{center}
\includegraphics[width=0.8\textwidth]{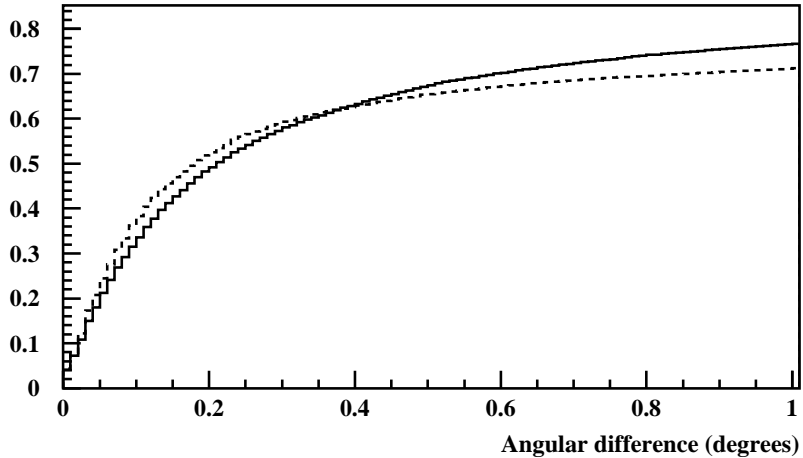}
\end{center}
\caption{Point spread function for the reconstructed events with 
at least three signal L1 hits (solid line). The corresponding point spread function 
for the alternative fitting technique used in HOURS is also shown (dashed line).
The y-axis shows the fraction of events with an angular difference less than the given value.}
\label{psf}
\end{figure}
Fig.~\ref{psf} presents the point spread function for reconstructed events, containing
at least three signal L1 hits in comparison with the alternative fitting technique used in HOURS.
The point spread function is the cumulative distribution function of the angular difference between the 
incident neutrino direction and the reconstructed muon direction.
The two techniques, for $E^{-2}$ neutrino spectrum, exhibit the same resolution as is presented in Fig.~\ref{psf} 
(the point spread functions have the same median of about $0.2^\circ$).
\section{Estimation of fake signal}
\label{fake}
In principle, taking into account the known direction of the source could produce fake signals.
In order to estimate the probability of such a fake signal production we have simulated background events 
(atmospheric neutrinos and atmospheric muon bundles) assuming
to be produced by a hypothetical point neutrino source with a random position isotropically distributed in a $4\pi$ solid angle.
We applied on the signal produced by these background tracks the filtering, prefit and track reconstruction procedures described in Sections \ref{filtering}, \ref{tripre} and \ref{reco}.

\begin{figure}
\begin{center}
\includegraphics[width=0.8\textwidth]{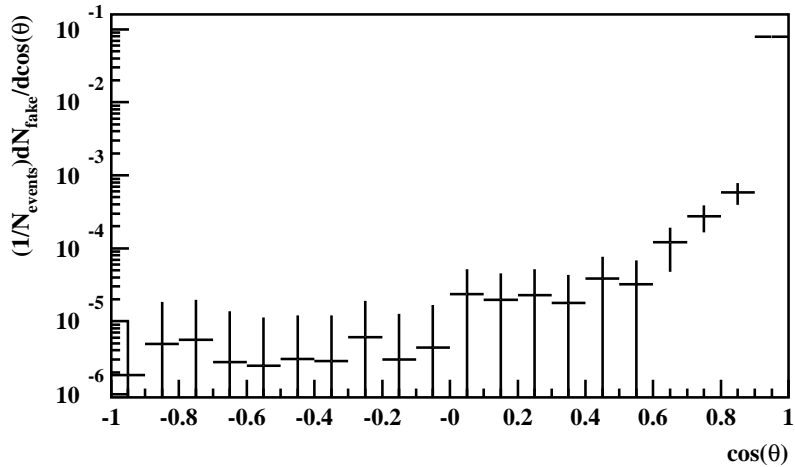}
\end{center}
\caption{Probability of an atmospheric neutrino induced event to produce fake signal versus the 
cosine of the angular difference, $\theta$, between the true neutrino direction and the 
assumed direction of the source.}
\label{faketracks}
\end{figure}
Fig.~\ref{faketracks} presents the probability of an atmospheric neutrino induced event to be reconstructed 
so that the angle between the reconstructed track and the assumed direction of the source is less than $1^\circ$.
This probability is shown as a function of the cosine of the angular difference, $\theta$, 
between the true atmospheric neutrino direction and the assumed direction of the source. 
The probability to produce a fake signal is negligible for $cos(\theta)<0.9$, 
i.e. atmospheric neutrino background has less than $10^{-3}$ probability to produce fake signal 
when the incident atmospheric neutrino comes from a direction which differs more than $25^\circ$ from the point source.

However, this fake signal can be further reduced by applying adequate tracking quality criteria.
Fake tracks have a low tracking quality which is reflected on the estimated tracking error, 
i.e. fake tracks carry a very small weight in the fit technique \cite{unbinned}.
\section{Conclusions}
We developed a triggering, filtering and prefit algorithm taking into account the apriori known
direction of the neutrino source. 
The performance of the KM3NeT detector in detecting/discovering point-like neutrino sources using 
this technique is under further evaluation. 
The performance of the combination of the known source direction technique with the fit technique enhances the 
discovery potential of the detector (a publication is under preparation).
\section*{Acknowledgments}
The KM3NeT project is supported by the EU in FP6 under Contract no. 011937
and in FP7 under Grant no. 212525
\bibliographystyle{elsarticle-num}

\end{document}